\documentclass[letterpaper,10pt,conference]{ieeeconf}  

\IEEEoverridecommandlockouts                              
\usepackage{cite}

\usepackage[dvips]{graphicx}
\usepackage[english]{babel}
\usepackage[latin1]{inputenc}
\usepackage[leqno]{amsmath}
\usepackage{amssymb}
\usepackage{subfig}
\usepackage{url}

\usepackage{cite} 
\usepackage{stfloats}

\newtheorem{theorem}{Theorem}
\newtheorem{lemma}{Lemma}

\title{\LARGE \bf
Quickest detection in coupled systems
}

\author{Olympia Hadjiliadis, Tobias Schaefer and H. Vincent Poor
\thanks{This work was supported by the RF-CUNY Collaborative grant 80209-04 15, the NSA-MSP grant 081103 and the NSF-DMS grant 0807396.}
\thanks{Olympia Hadjiliadis is with the Department of Mathematics, Brooklyn College, City University of New York,
and with the Departments of Computer Science and Mathematics, Graduate Center, City University of New York
        {\tt\small ohadjiliadis@brooklyn.cuny.edu}}
\thanks{Tobias Schaefer is with the Department of Mathematics, College of Staten Island, City University of New York, and with the Department of Physics, Graduate Center, City University of New York
        {\tt\small tobias@math.csi.cuny.edu}}
\thanks{H. Vincent Poor is with the Department of Electrical Engineering, Princeton University,
        {\tt\small poor@princeton.edu}}
}

\begin{document}

\maketitle
\thispagestyle{empty}
\pagestyle{empty}

\begin{abstract}

This work considers the problem of quickest
detection of signals in a coupled system of $N$ sensors, which receive
continuous sequential observations from the environment.
It is assumed that the signals, which are modeled a general
It\^{o} processes, are coupled across sensors, but that
their onset times may differ from sensor to sensor.  The objective
is the optimal detection of the first time at
which any sensor in the system receives a signal. The problem
is formulated as a stochastic optimization problem in which
an extended average Kullback-Leibler divergence criterion
is used as a measure of detection delay, with a constraint on
the mean time between false alarms. The case in
which the sensors employ cumulative sum (CUSUM) strategies
is considered, and it is proved that the minimum of $N$ CUSUMs
is asymptotically optimal as the mean time between false alarms
increases without bound.

\end{abstract}

\vspace*{2ex} \noindent {\bf Keywords: Kullback-Leibler divergence, CUSUM, quickest detection}

\section{INTRODUCTION}

 We are interested in the
problem of quickest detection of the onset of a signal in a system of $N$ sensors. We consider the situation in which, although
the observations in one sensor can affect the observations in another, the
onset of a signal can occur at different times (i.e., change points) in each of the $N$ sensors;
that is, the change points differ from sensor to sensor. As an example in which this situation arises
consider a system of sensors monitoring the health of a physical structure in which fault conditions are manifested by vibrations in the structure. Before a change
affects a given sensor, we have only noise in that sensor. Then, after a change, the system is vibrating and
thus the signal received in any location reflects a vibrating system.
Thus, observations at any given sensor are coupled with those received in other locations.
The change points observed at different sensors can occur at different times because the source of the vibrations
(i.e., the excitation) may arrive at different structural elements at different times. Relevant literature related to such models includes, for example,  \cite{BassAbdeBenv,BassBenvGourMeve,BassMeveGour,Ewin,HeylLammSas,Juan,PeetRoec}.

We assume that the probability law
of the observations is the same across sensors. This assumption although seemingly restrictive, is realistic
in view of the fact the system of sensors is coupled. We model the signals through continuous-time It\^o processes.
The advantage of such models is the fact that they can capture complex dependencies in the observations.
For example, an autoregressive process is a special case of the discrete-time equivalent of an Ornstein-Uhlenbeck process, which in turn, is a special case of an It\^o process.
Other special cases of this model include Markovian models, and
linear state-space systems commonly used in vibration-based structural analysis and health monitoring problems \cite{BassAbdeBenv,BassBenvGourMeve,BassMeveGour,Ewin,HeylLammSas,Juan,PeetRoec}. It is important to stress that
the fact that the system of $N$ sensors is coupled makes the probabilistic treatment of the problem equivalent to the one in which all observations become available in one location. The reason is that one integrated information flow is sufficient for describing such a system.

Our objective is to detect the first onset of a signal in such a system.  So far in the literature
of this type of problem (see \cite{Mous06,TartKim06,TartVeer03,TartVeer04,TartVeer08}) it has
been assumed that the change points are the same across sensors. Recently the case was also considered
of change points that propagate in a sensor array \cite{RaghVeer}. However, in this configuration the
propagation of the change points depends on the \underline{unknown} identity of the first sensor affected
and considers a restricted Markovian mechanism of propagation of the change.

In
this paper we consider the case in which the change points can be different and do not propagate
in any specific configuration. The objective is to detect the minimum (i.e., the first) of the change points.
We demonstrate that, in
the situation described above, at least asymptotically, the minimum of $N$
CUSUMs is asymptotically optimal in detecting the minimum of the $N$
different change points, as the mean time between false alarms tends
to $\infty$, with respect to an appropriately extended Kullback-Leibler divergence criterion
criterion \cite{Mous04} that incorporates the possibility of $N$
different change points.

In the next section we formulate the problem, discuss special cases of our It\^o models and demonstrate asymptotic optimality
(as the mean time between false alarms tends to $\infty$), in an extended min-max
Kullback-Leibler sense, of the minimum of $N$ CUSUM stopping times.
We finally discuss extensions of these results to the case of
different structures of observations in each sensor.

\section{FORMULATIONS \& RESULTS}

We sequentially observe the processes
$\{Z_t^{(i)};t \ge 0\}$ for all $i=1,\ldots,N$.
In order to formalize this problem we consider the measurable space $(\Omega, \mathcal{F})$, where
$\Omega = C[0,\infty]^N$ and $\mathcal{F}=\cup_{t >0} \mathcal{F}_t$ with
$\mathcal{F}_t=\sigma\{s \le t; Z_s^{(1)},\ldots, Z_s^{(N)}\}$.

The processes
$\{Z_t^{(i)};t \ge 0\}$ for all $i=1,\ldots,N$ are assumed to have the following
dynamics:
\begin{eqnarray}
\label{Itodynamics} dZ_t^{(i)} = \left \{
\begin{array}{ll}
dw_t^{(i)}  & t \le \tau_i \\
\alpha_t^{(i)}\,dt + dw_t^{(i)}&  t > \tau_i,
\end{array}
\right.
\end{eqnarray}

\noindent where $\{\alpha_t^{(i)};t \ge 0\}$ is a process on the same probability space adapted to the filtration $\{\mathcal{F}_t\}$ and $\{w_t^{(i)}; t\ge 0\}$ are independent standard Brownian motion. The case considered in this paper that in which $\alpha_t^{(i)}$ is the same for all $i$.
This can be described as a signal symmetry across sensors.

 We notice that $\{\mathcal{F}_t\}$ is the filtration generated by the observations received by all sensors. Thus by requiring that
$\alpha_t^{(i)}$ be $\mathcal{F}_t$-measurable for all $i$ , we have managed to capture the coupled nature of the system. In particular, in the special case in which, say,  $\alpha_t^{(1)}=-r\sum_{i=1}^N Z_t^{(i)}$,  (\ref{Itodynamics}) describes a process which displays an autoregressive (or its continuous equivalent \cite{Novi}) behavior in  $\{Z_t^{(1)};t \ge 0\}$, while still being coupled with the observations received by the other sensors. More specifically, the magnitude of each increment of the process $\{Z_t^{(1)};t \ge 0\}$ at each instant $t$ is not only affected by $Z_t^{(1)}$ but also by $Z_t^{(i)}$, $i=2,\ldots,N$ the observations at sensor $2,\ldots,N$. This couples the observations received in sensor $1$ with those received in sensors $2,\ldots,N$ at each instant $t$ and results in a system of interdependent sensors. We notice that the special case described above can also be written in the form of a linear state-space model as follows:
\begin{eqnarray*}
d \left( \begin{array}{l} Z_t^{(1)} \\ \ldots \\ Z_t^{(N)} \end{array} \right) & = & -r\left(\begin{array}{lcl} 1 & \ldots & 1 \\ \ldots & \ldots & \ldots \\ 1 & \ldots & 1 \end{array}\right) \left( \begin{array}{l} Z_t^{(1)} \\ \ldots \\ Z_t^{(N)} \end{array} \right) dt \\
& + &  I \left(\begin{array}{l} dW_t^{(1)} \\ \ldots \\ dW_t^{(N)} \end{array}\right).
\end{eqnarray*}
Autoregressive models and, more generally, linear state space models have been used to capture seismic signals, navigation systems, vibrating mechanical systems, etc. (see, e.g., \cite{BassNiki}). Another special case of (\ref{Itodynamics}) is
\begin{eqnarray*}
d \left( \begin{array}{l} Z_t^{(1)} \\ Z_t^{(2)} \end{array} \right) & = & \left(\begin{array}{ll} 0 &  1 \\ -1 & 0 \end{array}\right) dt+  \left(\begin{array}{l} dW_t^{(1)} \\ dW_t^{(2)} \end{array} \right),
\end{eqnarray*}
a model that describes sinusoidal waves driven by noise. Such a model could also be used to capture vibrating mechanical systems. The generality of \ref{Itodynamics} however is much greater than the special cases described above. This is seen in the fact that $\alpha_t^{(i)}$ at each instant $t$ can depend on the totality of the observed paths of each of the signals received up to time $t$.

On the space $\Omega$, we have the following family of probability measures
$\{P_{\tau_1,\ldots,\tau_N}\}$, where $P_{\tau_1,\ldots,\tau_N}$
corresponds to the measure generated on $\Omega$ by the processes
$(Z_t^{(1)},\ldots,Z_t^{(N)})$ when the change in the $N$-tuple
process occurs at time point $\tau_i$, $i=1,\ldots,N$. Notice that
the measure $P_{\infty,\ldots,\infty}$ corresponds to the measure
generated on $\Omega$ by $N$ independent standard Brownian motions.

Our objective is to find a stopping rule $T$ that balances the
trade-off between a small detection delay subject to a lower bound
on the mean-time between false alarms and will ultimately detect
$\min\{\tau_1,\ldots,\tau_N\}$. In what follows we will use
$\tilde{\tau}$ to denote
$\min\{\tau_1,\ldots,\tau_N\}$.

To this effect we propose a generalization of the $J_{KL}$ of \cite{Mous04}, namely

\begin{eqnarray} \label{JKL} \nonumber J_{KL}^{(N)}(T) = ~~~~~~~~~~~~~~~~~~~~~~~~~~~~~~~~~~~~~~~~~~~~~~~~~~~~~~\\ \nonumber \sup_{\tau_1,\ldots,\tau_N} \rm{essup}~E_{\tau_1,\ldots,\tau_N}\left\{\frac{1}{2}\left(\frac{1}{N}\int_{\tau_i}^T
 \sum_{i=1}^N(\alpha_s^{(i)})^2ds\right)~|~\mathcal{F}_{\tilde{\tau}}\right\}, \\
 \end{eqnarray}
where the supremum over $\tau_1, \ldots, \tau_N$ is taken over
the set in which $\min\{\tau_1,\ldots,\tau_N\} <\infty$. That is, we consider the
worst detection delay over all possible realizations of paths of the $N$-tuple of
stochastic processes $(Z_t^{(1)},\ldots,Z_t^{(N)})$ up to
$\min\{\tau_1,\ldots,\tau_N\}$ and then consider the worst detection delay over all
possible $N$-tuples $\{\tau_1,\ldots,\tau_N\}$ over a set in which at least one of
them is forced to take a finite value. This is because $T$ is a stopping rule meant
to detect the minimum of the $N$ change points and therefore if one of the $N$
processes undergoes a regime change, any unit of time by which $T$ delays in
reacting, should be counted towards the detection delay.
This gives rise to the
following stochastic optimization problem:
\begin{equation}
\begin{array}{c}
\displaystyle\inf_{T}J_{KL}^{(N)}(T),~~\textrm{subject to}
\\
E_{\infty,\ldots,\infty} \left\{\frac{1}{2}\int_0^T \frac{1}{N}\sum_{i=1}^N(\alpha_s^{(i)})^2 ds \right\} \geq \gamma.
\end{array}\label{eqnproblemKLG}
\end{equation}

\noindent The criterion in (\ref{JKL}) can be similarly motivated by considering the average over all sensors of the Kullback-Leibler divergence:
\begin{eqnarray}
\label{KLG} \nonumber
E_{\tau_1,\ldots,\tau_N}\left\{\left.\frac{1}{N}\left.\log\frac{dP_{\tau_1,\ldots,\tau_N}}{dP_{\infty,\ldots,\infty}}\right|_{\mathcal{F}_t}\right|\mathcal{F}_{\tilde{\tau} }\right\} \\
 = E_{\tau_1,\ldots,\tau_N}\left\{\frac{1}{N}\sum_{i=1}^N\int_{\tau_i}^t\frac{1}{2} (\alpha_r^{(i)})^2 dr |\mathcal{F}_{\tilde{\tau}}\right\}
\end{eqnarray}
where the last equality follows as long as
\begin{eqnarray} \label{eq:cond_6G}E_{\tau_1,\ldots,\tau_N}\left\{\int_{\tilde{\tau}}^t\left.(\alpha_r^{(i)})^2dr\right|\mathcal{F}_{\tilde{\tau}}\right\}<\infty~~a.s.
\end{eqnarray}
for all $i=1,\ldots,N$ and all $t<\infty$.

Using an argument similar to the randomization argument of \cite{HadjZhanPoor}, it is also possible to show that the optimal stopping rule $T^*$ must be an equalizer rule in that it would react at exactly the same time regardless of which change takes place first. In order to demonstrate this fact we begin by noting that minimization of
(\ref{JKL}) is equivalent to minimizing
\begin{eqnarray}
\label{eq_minim} \nonumber
\sup_{\tau_1,\ldots,\tau_N} \rm{essup}~E_{\tau_1,\ldots,\tau_N}\left\{\frac{1}{2}\left(\int_{\tilde{\tau}}^T \frac{1}{N} \sum_{i=1}^N
 (\alpha_s^{(i)})^2ds\right)~|~\mathcal{F}_{\tilde{\tau}}\right\} \\
 \end{eqnarray}

Now define
\begin{eqnarray*}
J_i^{(N)}(T)=~~~~~~~~~~~~~~~~~~~~~~~~~~~~~~~~~~~~~~~~~~~~~~~ \\
\sup_{\tau_i\le\tau_j,j\neq
i}\textrm{essup}E_{\tau_1,\ldots,\tau_N}\left\{\left.\left(\frac{1}{2}\int_{\tau_i}^T \frac{1}{N}\sum_{i=1}^N
 (\alpha_s^{(i)})^2ds\right)\right|\mathcal{F}_t\right\},
\end{eqnarray*}
for $i=1,\ldots,N$. That is, $J_i^{(N)}(T)$ is the detection delay
of the stopping rule $T$ when $\tau_i\le\min_{j\neq i}\{\tau_j\}$.
Then
\begin{eqnarray}\label{decomp}
\nonumber J_{KL}^{(N)}(T)=\max\left\{J_1^{(N)}(T),J_2^{(N)}(T),\ldots,J_N^{(N)}(T)\right\}.
\\
\end{eqnarray}
The optimal solution to (\ref{eqnproblemKLG}), $T^*$, satisfies
\begin{eqnarray}\label{A}
J_1^{(N)}(T^*)~=~J_2^{(N)}(T^*)~=~\ldots~=~J_N^{(N)}(T^*).
\end{eqnarray}
To see this, let us consider the case when $N=2$. Let $T$ be a
stopping rule such that $J_1^{(2)}(T)<J_2^{(2)}(T)$. Consider
another stopping rule $S$, which stops as $T$ does, but observes
$Z_t^{(2)}$ in place of $Z_t^{(1)}$ and $Z_t^{(1)}$ in place
of $Z_t^{(2)}$. It follows that
\begin{eqnarray*}
J_1^{(2)}(S)~=~J_2^{(2)}(T)~&\textrm{and}&~
J_2^{(2)}(S)~=~J_1^{(2)}(T).
\end{eqnarray*}
We trivially also have that
\begin{eqnarray*}
E_{\infty,\infty}\{S\}&=&E_{\infty,\infty}\{T\}.
\end{eqnarray*}
Now let us use a binary random variable $X\in\{0,1\}$, which is
independent of $\{\mathcal{F}_t\}$, to construct a randomized
stopping rule adapted to
$\hat{\mathcal{F}}_t=\mathcal{F}_t\vee\sigma(X)$,
\begin{eqnarray}\label{That}
\hat{T}&=&XT+(1-X)S.
\end{eqnarray}
It is easy to observe that
\begin{eqnarray*}
E_{\infty,\infty}\{\hat{T}\}&=&E_{\infty,\infty}\{T\},
\end{eqnarray*}
and
\begin{eqnarray*}
J_1^{(2)}(\hat{T})~=~J_2^{(2)}(\hat{T})~=~\frac{1}{2}\left[J_1^{(2)}(T)+J_2^{(2)}(T)\right]~<~J_2^{(2)}(T),
\end{eqnarray*}
which implies
\begin{eqnarray*}
J^{(2)}(\hat{T})&<&J^{(2)}(T),
\end{eqnarray*}
by (\ref{decomp}). Therefore the optimal solution to
(\ref{eqnproblemKLG}) must satisfy (\ref{A})\footnote{Although
$\hat{T}$ of equation (\ref{That}) is measurable with respect to the
enlarged filtration $\left\{\hat{\mathcal{F}}_t\right\}$, the optimal solution to
(\ref{eqnproblemKLG}) must be adapted to the original filtration
$\left\{\mathcal{F}_t\right\}$.}.

For a fixed $i$, and the dynamics of (\ref{Itodynamics}) the CUSUM stopping rule is
\begin{eqnarray}
\label{CUSUMIto}
T_\nu & = & \inf\{t \ge 0; y_t^{(i)}=\nu\},
\end{eqnarray}
where
\begin{eqnarray}\label{CUSUMstatIto}
y_t^{(i)} & = & u_t^{(i)}-m_t^{(i)},~i=1,\ldots,N
\end{eqnarray}
with
$m_t^{(i)}=\inf_{s \le t} u_s^{(i)}$, $i=1,\ldots,N$
and
\begin{eqnarray}
u_t^{(i)} & = &  \int_0^t \alpha_s^{(i)} dZ_s^{(i)}-\frac{1}{2} \int_0^t (\alpha_s^{(i)})^2 ds.
\end{eqnarray}

\noindent In the case that $N=1$, in which the drift denoted by $\alpha_t$ is measurable with respect to the filtration
generated by only one process, say $\{Z_t;t \ge 0\}$ the CUSUM stopping rule (\ref{CUSUMIto}) is optimal in minimizing the Kullback-Leibler divergence criterion of \cite{Mous04} subject to the false alarm constraint $E_\infty\{\frac{1}{2}\int_0^{T_\nu} \alpha_t^2 dt\} \ge \gamma$.
The $\nu$ in (\ref{CUSUMIto}) is chosen so that $E_\infty\left\{\frac{1}{2}\int_0^{T_{\nu}}\alpha_t^2dt \right\}\equiv
f(\nu)=\gamma$, with $f(\nu)=e^{\nu}-\nu-1$ (see
\cite{Mous04}) and
\begin{eqnarray}
\label{DD} J_{KL}^{(1)}(T_{\nu})\equiv E_0\left\{\frac{1}{2}\int_0^{T_{\nu}}\alpha_t^2dt\right\}=
f(-\nu).
\end{eqnarray}
The fact that the worst detection delay is the same as that incurred in the case in
which the change point is exactly $0$ is a consequence of the non-negativity of the
CUSUM process, from which it follows that the worst detection delay occurs when the
CUSUM process at the time of the change is at $0$ \cite{Mous04}.

The CUSUM stopping rule (\ref{CUSUMIto}) is an optimal solution to one-dimensional problem of detecting one change-point in the one-dimensional equivalent of (\ref{eqnproblemKLG}). The details can be found in \cite{Mous04} and \cite{PoorHadj}. It is important however to point out that a vital assumption necessary for the optimality of the CUSUM (\ref{CUSUMIto}) in \cite{Mous04} is
\begin{eqnarray}\nonumber
\label{Energy} P_{\tau_i}\left(\int_0^\infty {\alpha_s}^2
ds=\infty\right) & = & P_\infty\left(\int_0^\infty {\alpha_s}^2 ds =\infty\right) \\
& = & 1.
\end{eqnarray}
This assumption ensures the a.s. finiteness of the CUSUM stopping time (see \cite{LiptShir}), whose physical interpretation is that the
signal received after the change point has sufficient energy. We will thus assume that conditions (\ref{Energy}) are satisfied for all processes $\{\alpha_s^{(i)}\}$.

We remark here that if the $N$ change points were the same then the problem
(\ref{eqnproblemKLG}) is equivalent to observing only one stochastic process which is
now $N$-dimensional. Thus, in this case, the detection delay and mean time between
false alarms are given by the formulas in the above paragraph.

Returning to problem (\ref{eqnproblemKLG}), it is easily seen that in seeking solutions
to this problem, we can restrict our attention to stopping times that achieve the
false alarm constraint with equality \cite{Mous86}. The optimality of the CUSUM
stopping rule in the presence of only one observation process suggests that a CUSUM
type of stopping rule might display similar optimality properties in the case of
multiple observation processes. In particular, an intuitively appealing rule, when
the detection of $\min\{\tau_1,\ldots,\tau_N\}$ is of interest, is $T_{h}=T_{h}^1
\wedge \ldots \wedge T_{h}^N$, where $T_h^i$ is the CUSUM stopping rule for the
process $\{Z_t^{(i)};t \ge 0\}$ for $i=1,\ldots,N$. That is, we use what is known
as a multi-chart CUSUM stopping time \cite{Tart05}, which can be written as
\begin{eqnarray}
\label{CUSUMmultichart} T_{h} & = & \inf\left\{t \ge 0;
\max\{y_t^{(1)},\ldots,y_t^{(N)}\} \ge h\right\},
\end{eqnarray}

\noindent where $$y_t^{(i)}=\sup_{0 \le \tau_i \le t}\left.\log
\frac{dP_{\tau_i}}{dP_\infty}\right|_{\mathcal{F}_t},$$

\noindent and the $P_{\tau_i}$ are the restrictions of the measure
$P_{\tau_1,\ldots,\tau_N}$ to $C[0,\infty)$.

It is easy to see that (\ref{CUSUMmultichart}) is an equalizer rule. That is, it satisfies (\ref{A}).
This follows from the assumption that $\{\alpha_t^{(i)}\}$ are the same for all $i$.

Moreover,
\begin{eqnarray}
\label{eq} \nonumber
J_{KL}^{(N)}(T_{h})& = & E_{0,\infty,\ldots,\infty}\left\{\frac{1}{2}\int_0^{T_{h}}(\alpha_t^{(1)})^2dt\right\} \\
\nonumber & = &
E_{\infty,0,\infty,\ldots,\infty}\left\{\frac{1}{2}\int_0^{T_{h}}(\alpha_t^{(1)})^2dt\right\}
\\ \nonumber & = & \ldots \\& = & E_{\infty,\ldots,\infty,0}\left\{\frac{1}{2}\int_0^{T_{h}}(\alpha_t^{(1)})^2dt\right\}.
\end{eqnarray}

This is because the worst detection delay occurs when at least one of the $N$
processes does not change regime. Thus, the worst detection delay will
occur when none of the other processes changes regime and due to the non-negativity
of the CUSUM process the worst detection delay will occur when the remaining one
processes is exactly at $0$.

Notice that the threshold $h$ is used for the multi-chart CUSUM stopping rule
(\ref{CUSUMmultichart}) in order to distinguish it from $\nu$ the threshold used for
the one sided CUSUM stopping rule (\ref{CUSUMIto}).

In what follows we will demonstrate asymptotic optimality of (\ref{CUSUMmultichart})
as $\gamma \to \infty$. In view of the constraint in (\ref{eqnproblemKLG}), the assumption that $\{\alpha_t^{(i)}\}$ are the same for all $i$ and (\ref{eq}), in
order to assess the optimality properties of the multi-chart CUSUM rule
(\ref{CUSUMmultichart}), we will need to begin by evaluating
$E_{0,\infty,\ldots,\infty}\left\{\frac{1}{2}\int_0^{T_h} (\alpha_t^{(1)})^2dt\right\}$ and
$E_{\infty,\ldots,\infty}\left\{\frac{1}{2}\int_0^{T_h} (\alpha_t^{(1)})^2dt\right\}$.

In order to demonstrate asymptotic optimality of (\ref{CUSUMmultichart}) we bound
the detection delay $J_{KL}^{(N)}$ of the unknown optimal stopping rule $T^*$ by
\begin{eqnarray}\label{UB}
E_{0,\infty,\ldots,\infty}\left\{\frac{1}{2}\int_0^{T_h}(\alpha_t^{(1)})^2dt\right\} & > & J_{KL}^{(N)}(T^*),
\end{eqnarray}
where $h$ is chosen so that
\begin{eqnarray}\label{FACSh} E_{\infty,\ldots, \infty}\left\{\frac{1}{2}\int_0^{T_h} (\alpha_t^{(1)})^2dt\right\}& = & \gamma.\end{eqnarray}
It is also obvious that $J_{KL}^{(N)}(T^*)$ is bounded from below by the detection
delay of the one CUSUM when there is only one observation process, in view of the
fact that
\begin{eqnarray}\label{ineq} \nonumber
& \sup_{\tau_1,\ldots,\tau_N}\textrm{essup}E_{\tau_1,\ldots,\tau_N}
\left\{\frac{1}{2}\int_{\tilde{\tau}}^{T} (\alpha_t^{(1)})^2dt | \mathcal{F}_{\tilde{\tau}} \right\} \ge &
\\ & \ge \sup_{\tau_1}\textrm{essup}\,E_{\tau_1} \left\{\frac{1}{2}\int_{\tau_1}^{T} \alpha_t^2dt |
\mathcal{F}_{\tau_1}^{(1)} \right\}, &
\end{eqnarray}
where $\alpha_t$ is measurable w.r.t. the filtration generated by the $1$-dimensional process $\{Z_t^{(1)}\}$, denoted by $\{\mathcal{F}_t^{(1)}\}$,
and is the projection of $\{\alpha_t^{(1)}\}$ on the filtration $\{\mathcal{F}_t^{(1)}\}$. 

The stopping time that minimizes the right hand side is the CUSUM stopping rule
$T_\nu$ of (\ref{CUSUMIto}), with $\nu$ chosen so as to satisfy
\begin{eqnarray}\label{FACSnu} E_\infty\left\{\frac{1}{2}\int_0^{T_\nu}\alpha_t^2 dt\right\}& = & \gamma.\end{eqnarray}
We will demonstrate that the difference between the
upper and lower bounds
\begin{eqnarray}
\label{ULB} \nonumber E_{0,\infty,\ldots,\infty}\left\{\frac{1}{2}\int_0^{T_h}(\alpha_t^{(1)})^2dt \right\} & >  &J_{KL}^{(N)}(T^*) \\ \nonumber & > &
E_0\left\{\frac{1}{2}\int_0^{T_\nu}\alpha_t^2dt\right\}, \\
& ~ &
\end{eqnarray}
is bounded by a constant as $\gamma \to \infty$, with $h$ and $\nu$
satisfying (\ref{FACSh}) and (\ref{FACSnu}), respectively.

\begin{lemma}
\label{asymptotics} Suppose that $\{\alpha_t^{(i)}\}$ are the same for all $i$. We have
\begin{eqnarray}\label{AsymptoticSh}  \nonumber  E_{0,\infty,\ldots,\infty}\left\{\frac{1}{2}\int_0^{T_h} (\alpha_t^{(1)})^2 dt \right\} =  \left[\log \gamma +\log
N-1 + o(1)\right] ,\\
\end{eqnarray} as $\gamma \to \infty$
\end{lemma}
{\bf Proof:} Please refer to the Appendix for a sketch of the proof.
Moreover, it is easily seen from (\ref{DD}) that
\begin{eqnarray}\label{AsymptoticSn} E_0\left\{\frac{1}{2}\int_0^{T_\nu}\alpha_t^2 dt\right\} & = & \left[\log
\gamma  -1+o(1)\right].
\end{eqnarray}
Thus we have the following result.
\begin{theorem}
\label{main} Suppose that $\{\alpha_t^{(i)}\}$ are the same for all $i$. Then the difference in detection delay $J_{KL}^{(N)}$ of the
unknown optimal stopping rule $T^*$ and the detection delay of $T_h$
of (\ref{CUSUMmultichart}) with $h$ satisfying (\ref{FACSh}) is
bounded above by $$\log N,$$ as $\gamma\to \infty$.
\end{theorem}
{\bf Proof:} The proof follows from Lemma \ref{asymptotics} and
(\ref{AsymptoticSn}).
\vspace*{2ex}

\noindent {\it { Remark:}} Since
$J_{KL}^{(N)}(T_h)$ increases without bound as $\gamma \to \infty$,
Theorem \ref{main} asserts the asymptotic optimality of $T_h$.

\section{CONCLUSIONS AND FUTURE WORKS}

In this paper we have demonstrated the asymptotic optimality of the minimum of $N$ CUSUMs for detecting the minimum of $N$ different change points in a coupled system of $N$ sensors which receive sequential observations from the environment. We have allowed for a general dependence structure in the observations and we have shown that the $N$-CUSUM stopping rule is asymptotically optimal, as the mean time to the first false alarm increases without bound, in detecting the minimum of $N$ different change-points in the sense that it minimizes a worst average Kullback-Leibler divergence criterion. This has been seen by the fact that the difference in detection delay of the proposed $N$-CUSUM stopping rule and the unknown optimal stopping rule is bounded above by the constant $\log N$.
An interesting extension of this work would incorporate the fact that the distributions of the signals received in different sensors may be different. In this case the fact that the optimal stopping rule has to be an equalizer rule (i.e. satisfy (\ref{A})) would determine the optimal selection of thresholds in each sensor which in the general case should be different.

\section{ACKNOWLEDGMENTS}

The authors are grateful to Professor Jay Rosen for promoting this collaboration.


\section{Appendix}

As an illustration for the general case, let us prove the result for $N=2$.

We begin by deriving the Partial Differential equations satisfied by the functions
\begin{itemize}
\item $\tilde{S}(\tilde{x},\tilde{y})=E_{0,\infty}^{(\tilde{x},\tilde{y})}\left\{\frac{1}{2}\int_0^{T_h} (\alpha_t^{(1)})^2 dt
\right\},$
\item $\tilde{T}(\tilde{x},\tilde{y})=E_{\infty,\infty}^{(\tilde{x},\tilde{y})}\left\{\frac{1}{2}\int_0^{T_h}  (\alpha_t^{(1)})^2 dt  \right\}$,
\end{itemize}
where the subscript $(\tilde{x},\tilde{y})$ indicates the indicates the initial value of the pair of CUSUM processes
$(y_t^{(1)},y_t^{(2)})$. With this representation, it is easy to see that
$E_{0,\infty}\left\{\frac{1}{2}\int_0^{T_h} (\alpha_t^{(1)})^2 dt
\right\}=\tilde{S}(0,0)$ and $E_{\infty,\infty}\left\{\frac{1}{2}\int_0^{T_h} (\alpha_t^{(1)})^2 dt
\right\}=\tilde{T}(0,0)$. In the sequel we will denote by $\tilde{T}_{\tilde{x}}$, $\tilde{T}_{\tilde{y}}$, $\tilde{S}_{\tilde{x}}$, $\tilde{S}_{\tilde{y}}$, the first partial derivatives of $\tilde{T}$ and $\tilde{S}$
with respect to $\tilde{x}$ and $\tilde{y}$ respectively. Similarly, we will denote by $\tilde{T}_{\tilde{x}\tilde{x}}$, $\tilde{T}_{\tilde{y}\tilde{y}}$, $\tilde{S}_{\tilde{x}\tilde{x}}$, $\tilde{S}_{\tilde{y}\tilde{y}}$ the second partials.

Using It\^o's rule \cite{Okse}, we have
\begin{multline}
\tilde{T}(y_t^{(1)},y_t^{(2)})-\tilde{T}(\tilde{x},\tilde{y}) = \int_0^{t} \alpha_s^{(1)} \tilde{T}_{\tilde{x}}dw_s^{(1)}+\alpha_s^{(2)} \tilde{T}_{\tilde{y}}dw_s^{(2)} \\  -  \int_0^t   \tilde{T}_{\tilde{x}}(y_s^{(1)},y_s^{(2)})dm_s^{(1)}+ \tilde{T}_{\tilde{y}}(y_s^{(1)},y_s^{(2)})dm_s^{(2)} \\
 +  \int_0^t (\alpha_s^{(i)})^2(\tilde{T}_{\tilde{x}\tilde{x}}+\tilde{T}_{\tilde{y}\tilde{y}}-\tilde{T}_{\tilde{x}}-\tilde{T}_{\tilde{y}})ds,
\label{SDE}
\end{multline}
where the arguments of each of the above functions are $(\tilde{y}_s^{(1)},\tilde{y}_s^{(2)})$ when omitted and where in the last line we use the fact that $\alpha_s^{(i)}$ are of the same form for all $i$. Evaluating the above equation at $T_h$ and taking expectations under the $P_{\infty,\infty}$ measure, while using conditions (\ref{eq:cond_6G}), (\ref{Energy}), we obtain that $\tilde{T}$ has to satisfy
\begin{eqnarray}
\label{PDEF}
~~~~\tilde{T}_{\tilde{x}\tilde{x}}+\tilde{T}_{\tilde{y}\tilde{y}}-\tilde{T}_{\tilde{x}}-\tilde{T}_{\tilde{y}} & = & -1,~(\tilde{x},\tilde{y}) \in {\mathcal{\tilde{D}}}=[0,h]^2,
\end{eqnarray}
with the Dirichlet boundary conditions
\begin{equation} \label{dirBC}
\tilde{T}(\tilde{x},\tilde{y})|_{\tilde{x}=h} = \tilde{T}(\tilde{x},\tilde{y})|_{\tilde{y}=h}=0\,
\end{equation}
and the Neumann boundary conditions
\begin{equation} \label{neuBC}
\left. \frac{\partial \tilde{T}}{\partial \tilde{x}}\right|_{\tilde{x}=0}= \left. \frac{\partial \tilde{T}}{\partial \tilde{y}}\right|_{\tilde{y}=0}=0.
\end{equation}
Notice that the Neumann boundary conditions ensure that the terms in the second line of (\ref{SDE}) vanish. Similarly, $\tilde{S}$ satisfies
\begin{eqnarray}
\label{PDEG}
~~~~~~~\tilde{S}_{\tilde{x}\tilde{x}}+\tilde{S}_{\tilde{y}\tilde{y}}+\tilde{S}_{\tilde{x}}-\tilde{S}_{\tilde{y}} & = & -1,~(\tilde{x},\tilde{y}) \in {\mathcal{\tilde{D}}}=[0,h]^2,
\end{eqnarray}
with the same boundary conditions as $\tilde{T}$.

We can now introduce a change of variable $x=\frac{\tilde{x}}{h}$ and $y=\frac{\tilde{y}}{h}$. By setting $\epsilon=\frac{1}{h}$, we can rewrite (\ref{PDEF}) as
\begin{eqnarray}
\label{PDEtildeT} \\ \nonumber
\epsilon^2\tilde{T}_{xx}+\epsilon^2\tilde{T}_{yy}-\epsilon\tilde{T}_{x}-\epsilon\tilde{T}_{y} & = & -1,~(x,y) \in {\mathcal{D}}=[0,1]^2,
\end{eqnarray}
with the Dirichlet boundary conditions
\begin{equation} \label{dirBCn}
\tilde{T}(x,y)|_{x=1} = \tilde{T}(x,y)|_{y=1}=0\,
\end{equation}
and the Neumann boundary conditions (\ref{neuBC}). By letting $\epsilon \tilde{T}=T$, we now obtain
\begin{eqnarray}
\label{PDET} & ~ & \\ \nonumber
\epsilon T_{xx}+\epsilon T_{yy}- T_{x}-T_{y} & = & -1,~(x,y) \in {\mathcal{D}}=[0,1]^2,
\end{eqnarray}
with $T$ satisfying the Dirichlet boundary conditions of (\ref{dirBCn}) and the Neumann condition of (\ref{neuBC}).
We are interested in the asymptotics of $T(0,0)$ for small values of
$\epsilon$ (or equivalently large values of $h$). $T(0,0)$ can be interpreted as the mean exit time of a particle
that is placed initially at the origin, with reflecting boundaries
along the axes and absorbing boundaries on the top and the right side
of the rectangular domain ${\mathcal{D}}$.
In order to solve the above problem, we note, that we can write the
solution $T$ as
\begin{equation}
\label{integral}
T(x,y) = \int_0^{\infty}G(x,y,t)\,dt
\end{equation}
where $G$ denotes the probability that the particle, initially placed at a
point $(x,y)$ in ${\mathcal{D}}$ leaves the domain ${\mathcal{D}}$ at
a time $\tau>t$. The evolution of $G$ is then governed by the backward Fokker-Planck equation:
\begin{equation} \label{PDE_G}
\frac{\partial G}{\partial t} = \epsilon\Delta G -\frac{\partial G}{\partial x}
- \frac{\partial G}{\partial y}\,.
\end{equation}
Boundary conditions for $G$ correspond to boundary conditions of $T$ and
the initial condition of $G$ is given by the fact that, at $t=0$,
$G$ has the value $1$ in ${\mathcal{D}}$.

In the case of the particular geometry under consideration, we can find
an approximate solution to (\ref{PDE_G}) and use this to find $T$. This
 is due to the fact that, for a rectangular domain under the assumptions given,
the solution of (\ref{PDE_G}) can be found by simple separation of variables,
hence we find $G$ as a product of the form
\begin{equation}
\label{product}
G(x,y,t) = G_1(x,t)G_2(y,t),
\end{equation}
where $G_1$ satisfies the equation
\begin{equation} \label{PDE_G1}
\frac{\partial G_1}{\partial t} = \epsilon \frac{\partial^2G_1}{\partial x^2}-
\frac{\partial G_1}{\partial x}
\end{equation}
on $[0,1]$ with reflecting boundary at $0$ and absorbing boundary at $1$.
The same holds for $G_2$ with respect to the variable $y$.

In order to solve (\ref{PDE_G1}), we apply a Laplace transform in $t$ and
obtain for $\tilde G_1=\tilde G_1(s,x)$ the ordinary differential equation
\begin{equation} \label{BVPinG}
s\tilde G_1 -1 = \epsilon \tilde G_1''-\tilde G_1'.
\end{equation}
Making use of the fact that $\epsilon$ is small, we find as leading order approximation
to the solution of (\ref{BVPinG}):
\begin{equation}
\tilde G_1(0,s)\approx \frac{\epsilon\,{\mathrm{e}}^{1/\epsilon}}
                            {\epsilon s\,{\mathrm{e}}^{1/\epsilon}+1}.
\end{equation}
For this approximation it is simple to find the inverse Laplace transform to obtain
\begin{equation} \label{GtMinus}
G_1(0,t)\approx \exp\left(-\frac{1}{\epsilon}{\mathrm{e}}^{-1/\epsilon}t\right).
\end{equation}
Using this formula for both $G_1(0,t)$ and $G_2(0,t)$ we obtain immediately for
$T(0,0)$ in (\ref{PDET}) the asymptotic formula
\begin{equation}
\label{asymptapprox}
T(0,0) \approx \frac{1}{2}\epsilon\,{\mathrm{e}}^{1/\epsilon}\,
\end{equation}
from which it follows that $\tilde{T}(0,0)\approx \frac{1}{2}\,{\mathrm{e}}^{1/\epsilon}$. Setting
$\tilde{T}(0,0)=\gamma$, and using $h=\frac{1}{\epsilon}$, we further obtain that as $\gamma \to \infty$,
$h \approx \log \gamma +\log 2$.

For the asymptotic formula of $\tilde{S}(0,0)$ of (\ref{PDEG}), we also let $S=\epsilon\tilde{S}$ and use the same change of variable as in the previous case.
The only difference is that we have to solve for $\tilde G_1$ the different problem
\begin{equation}
\label{PDE+}
s\tilde G_1 -1 = \epsilon \tilde G_1''+\tilde G_1'.
\end{equation}
In this case, the approximate solution takes the form
\begin{equation}
\tilde G_1(0,s)\approx \frac{1-{\mathrm{e}}^{-s}}{s}-\epsilon\,{\mathrm{e}}^{-s}\,(1+s).
\end{equation}
From here we obtain after inverse Laplace transform
\begin{equation} \label{GtPlus}
G_1(0,t) \approx {\mathcal{H}}(1-t) - \epsilon\,\left(\delta(t-1)+\delta'(t-1)\right)\,,
\end{equation}
where ${\mathcal{H}}$ denotes the Heaviside function and $\delta$ denotes the Dirac delta distribution. Combining
the formulas (\ref{GtPlus}) for $G_1$ and (\ref{GtMinus}) for $G_2$ we find as approximation of $S(0,0)$ for
the problem (\ref{PDEG})
\begin{equation}
\label{integralS}
S(0,0)=\int_0^{\infty}G_1(0,t)G_2(0,t)\,dt \approx 1-\epsilon,
\end{equation}
from which we obtain $\tilde{S}(0,0) \approx \frac{1}{\epsilon}-1=h-1$, from which it follows that
$\tilde{S}(0,0) \approx \log \gamma+\log 2-1$ as $\gamma \to \infty$.

Using the same derivational steps it is possible to generalize to $N$ sensors.
In particular, in this case the integrand for $T(x_1,\ldots,x_N)$ in (\ref{integral})
becomes the product (see (\ref{product})) of $N$ functions, $G_1(x_1,t),\ldots,G_N(x_N,t)$
each of which satisfies equation (\ref{PDE_G1}) with the same boundary conditions with respect to their respective variables. Their respective Laplace transforms satisfy (\ref{BVPinG}).
This leads to
\begin{eqnarray}
\label{ASAPN}
T(0,\ldots,0) \approx \frac{1}{N}\epsilon\,{\mathrm{e}}^{1/\epsilon}.
\end{eqnarray}

Similarly, $S(0,\ldots,0)$ takes the form (\ref{integralS}), with integrand consisting of the product of $N$ functions, the Laplace transform of the first of which satisfies (\ref{PDE+}) and the Laplace transforms of the others satisfy (\ref{BVPinG}). Following the same steps as before, this leads to the asymptotic formula
\begin{equation}
\label{ASAPNS}
S(0,\ldots,0) \approx 1-\epsilon.
\end{equation}
Using (\ref{ASAPN}) and (\ref{ASAPNS}), we derive $\tilde{S}(0,0) \approx \log \gamma+\log N-1$ as $\gamma \to \infty$. This completes the proof of Lemma \ref{asymptotics}.

\end{document}